\documentclass[final,5p,times,twocolumn]{elsarticle}

\usepackage{lineno,hyperref}
\modulolinenumbers[5]

\journal{Journal of \LaTeX\ Templates}

\usepackage{graphicx}
\usepackage{dcolumn}
\usepackage{bm}
\usepackage{xcolor}

\usepackage{mathtools}

\DeclarePairedDelimiter\abs{\lvert}{\rvert}%
\DeclarePairedDelimiter\norm{\lVert}{\rVert}%

\usepackage{threeparttable}

\makeatletter
\let\oldabs\abs
\def\abs{\@ifstar{\oldabs}{\oldabs*}}
\let\oldnorm\norm
\def\norm{\@ifstar{\oldnorm}{\oldnorm*}}
\makeatother

\mathchardef\mhyphen="2D 

\newcommand{\beam}{\textrm{\fontsize{7}{8}\selectfont beam}}

\newcommand{\FWHM}{\textsc{FWHM}}

\makeatletter
\DeclareRobustCommand{\mypm}{\mathbin{\mathpalette\@mypm\relax}}
\DeclareRobustCommand{\@mypm}[2]{\ooalign{%
  \raisebox{.1\height}{$#1+$}\cr
  \smash{\raisebox{-.6\height}{$#1-$}}\cr}}
\makeatother

\definecolor{darkgrey}{rgb}{0.130416, 0.130416, 0.130416}

\begin{document}

\begin{frontmatter}

\title{Investigating the predicted breathing-mode excitation of the Hoyle state}


\author[iThembaLABS,UiO,Stb]{K.~C.~W.~Li}

\author[iThembaLABS]{F.~D.~Smit}%

\author[iThembaLABS,DepartmentOfPhysicsAndAstronomy_TAMU,CyclotronInstitute_TAMU,Wits]{P.~Adsley}%

\author[iThembaLABS]{R.~Neveling}%

\author[iThembaLABS,Stb]{P.~Papka}%

\author[Kurchatov,Flerov]{E.~Nikolskii}%

\author[iThembaLABS,Stb]{\\J.~W.~Br\"ummer}%

\author[iThembaLABS,Wits]{L.~M.~Donaldson}%

\author[Birmingham]{M.~Freer}%

\author[ESRIG]{M. N.~Harakeh}%


\author[iThembaLABS]{F.~Nemulodi}%

\author[iThembaLABS,Wits]{L.~Pellegri}%

\author[UWC,CIEMAT]{V.~Pesudo}%

\author[iThembaLABS,Wits]{M.~Wiedeking}%


\author[iThembaLABS]{\\\color{darkgrey} E.~Z.~Buthelezi}%

\author[Flerov]{\color{darkgrey} V.~Chudoba}%

\author[iThembaLABS]{\color{darkgrey} S.~V.~F{\"o}rtsch}%

\author[iThembaLABS]{\color{darkgrey} P.~Jones}%

\author[UWC]{\color{darkgrey} M.~Kamil}%

\author[iThembaLABS]{\color{darkgrey} J.~P.~Mira}%

\author[iThembaLABS,UWC]{\color{darkgrey} G.~G.~O'Neill}%

\author[Wits]{\color{darkgrey} E.~Sideras-Haddad}%

\author[UWC]{\color{darkgrey} B.~Singh}%

\author[UiO]{\color{darkgrey} S.~Siem}%

\author[iThembaLABS]{\color{darkgrey} G.~F.~Steyn}%

\author[iThembaLABS,Stb]{\color{darkgrey} J.~A.~Swartz}%

\author[iThembaLABS,Wits]{\color{darkgrey} I.~T.~Usman}%

\author[Stb]{\color{darkgrey} J.~J.~van Zyl}%

\address[iThembaLABS]{iThemba LABS, National Research Foundation, PO Box 722, Somerset West 7129, South Africa}
\address[UiO]{Department of Physics, University of Oslo, N-0316 Oslo, Norway}
\address[Stb]{Department of Physics, University of Stellenbosch, Private Bag X1, 7602 Matieland, South Africa}
\address[DepartmentOfPhysicsAndAstronomy_TAMU]{Department of Physics and Astronomy, Texas A\&M University, College Station, Texas 77843, USA}
\address[CyclotronInstitute_TAMU]{Cyclotron Institute, Texas A\&M University, College Station, Texas 77843, USA}
\address[Wits]{School of Physics, University of the Witwatersrand, Johannesburg 2050, South Africa}
\address[Kurchatov]{NRC Kurchatov Institute, Ru-123182 Moscow, Russia}
\address[Flerov]{Flerov Laboratory of Nuclear Reactions, JINR, RU-141980 Dubna, Russia}
\address[Birmingham]{School of Physics and Astronomy, University of Birmingham, Edgbaston, Birmingham, B15 2TT, United Kingdom}
\address[ESRIG]{Nuclear Energy Group, ESRIG, University of Groningen, 9747 AA Groningen, The Netherlands}
\address[UWC]{Department of Physics and Astronomy, University of the Western Cape, P/B X17, Bellville 7535, South Africa}
\address[CIEMAT]{Centro de Investigaciones Energ\'eticas, Medioambientales y Tecnol\'ogicas, Madrid 28040, Spain}




\begin{abstract}
Knowledge of the low-lying monopole strength in $\mathrm{^{12}C}$---the Hoyle state in particular---is crucial for our understanding of both the astrophysically important $3\alpha$ reaction and of $\alpha$-particle clustering. 
Multiple theoretical models have predicted a breathing mode of the Hoyle State at $E_{x} \approx 9$ MeV, corresponding to a radial in-phase oscillation of the underlying $\alpha$ clusters.
The $\mathrm{^{12}C}(\alpha, \alpha^{\prime})\mathrm{^{12}C}$ and $\mathrm{^{14}C}(p, t)\mathrm{^{12}C}$ reactions were employed to populate states in $^{12}$C in order to search for this predicted breathing mode. 
A self-consistent, simultaneous analysis of the inclusive spectra with \textbf{R}-matrix lineshapes, together with angular distributions of charged-particle decay, yielded clear evidence for excess monopole strength at $E_{x} \approx 9$ MeV which is highly collective. 
Reproduction of the experimentally observed inclusive yields using a fit, with consistent population ratios for the various broad states, required an additional source of monopole strength.
The interpretation of this additional monopole resonance as the breathing-mode excitation of the Hoyle state would provide evidence supporting a $\mathcal{D}_{3h}$ symmetry for the Hoyle state itself.
The excess monopole strength may complicate analysis of the properties of the Hoyle state, modifying the temperature dependence of the $3\alpha$ rate at $T_{9} \gtrsim 2$ and ultimately, the predicted nucleosynthesis in explosive stars.
\end{abstract}


\end{frontmatter}


The emergent phenomenon of $\alpha$-particle clustering in light nuclei has garnered significant interest for both nuclear structure and astrophysics, resulting in predictions such as dilute densities and Bose-Einstein condensation \cite{BoseSN,EinsteinA}.
Of particular interest is $\mathrm{^{12}C}$, which exhibits  shell-model and $\alpha$-cluster structures.
The Hoyle state in $^{12}$C is the archetypal $\alpha$-cluster state \cite{FREER20141}, remaining the focus of considerable study, including efforts to measure its direct \cite{PhysRevLett.113.102501, PhysRevLett.119.132501, PhysRevLett.119.132502, TKRana_2019, PhysRevC.101.021302, PhysRevC.102.041303}, $\gamma$ \cite{PhysRevLett.125.182701} and $E0$ decay branching ratios \cite{PhysRevC.102.024320}.
Such decay branches are astrophysically significant; the Hoyle state mediates the $3\alpha$ reaction producing $\mathrm{^{12}C}$.
The $3\alpha$ reaction rate requires an accurate description of the Hoyle-state properties (e.g., resonance energy $E_{r}$, partial and total widths, $\Gamma_{i}$ and $\Gamma$) and their evolution with excitation energy.
In the region $0.1 < T_{9} < 2$ ($T_{9} = T /10^{9}$ K), the $3\alpha$ rate is determined by the $Q$ value, and the $\gamma$-ray and pair-production partial widths \cite{HOUFynbo_2005}. 
At $T_{9} > 2$, higher-lying resonances, such as the rotational $2^{+}$ excitation of the Hoyle state \cite{PhysRevC.84.054308, PhysRevC.86.034320, PhysRevLett.110.152502}, and the Hoyle state's ``ghost'' contribute to the $3\alpha$ reaction rate \cite{HOUFynbo_2005}.
The ghost of the Hoyle state corresponds to a pronounced high-energy tail with a local maximum above the narrow primary peak of the Hoyle state.
This phenomenon results from the strong $\alpha$-cluster character of the Hoyle state and its proximity to the $\alpha$-separation energy \cite{FREER20141, 1976AuJPh_29_245B}.
Understanding the evolution of the Hoyle state's properties (e.g. partial widths and branching ratios) with excitation energy, and the number and nature of the additional excited levels above the Hoyle state, is vital in computing the $3\alpha$ rate in explosive burning.
The partial widths of the Hoyle state, and subsequently the $3\alpha$ reaction, may be inaccurately estimated if there are unaccounted-for sources of monopole strength or if the phenomenological description of the Hoyle state is insufficient, failing to properly describe the observed monopole strength.

Multiple independent predictions of the Hoyle-state breathing-mode monopole strength with a variety of theoretical models have been made in recent years \cite{PhysRevC.71.021301, CKurokawa, 10.1093/ptep/ptt048, PhysRevC.82.034307, PhysRevC.94.044319, PhysRevC.99.064327}.
This breathing-mode excitation corresponds to a radial in-phase oscillation of the underlying $\alpha$ clusters.
This excitation is predicted to lie at $E_{x} \approx 8-9.5$ MeV between two previously established sources of monopole strength: the $0_{2}^{+}$ Hoyle state at $E_{x} = 7.65407(19)$ MeV with its associated ghost and a broad $0_{3}^{+}$ state at $E_{x} \approx 10-11$ MeV \cite{HYLDEGAARD2009459,PhysRevC.80.044304,ENSDF} with $\Gamma \approx 3$ MeV \cite{ENSDF}.
Some models \cite{PhysRevLett.106.192501, PhysRevLett.109.252501, PhysRevC.94.044319, PhysRevC.82.034307} also predict an additional source of monopole strength above $E_x \approx 10$ MeV, which may correspond to the previously observed $0_{3}^{+}$ state.

A novel interpretation of the structure of $\mathrm{^{12}C}$ with $\mathcal{D}_{3h}$ symmetry, historically employed to study triatomic molecules, provided a compelling description of high-spin states as the rotational excitations of an equilateral triangular structure \cite{PhysRevC.61.067305, BIJKER2002334, PhysRevLett.113.012502}.
A breathing-mode excitation of a 3$\alpha$ cluster state, depicted as the radial in-phase oscillation of the component $\alpha$ clusters, corresponds to a characteristic vibrational mode of $\mathcal{D}_{3h}$ symmetry.
The existence of such a breathing-mode excitation, specifically built on the Hoyle state, would suggest $\mathcal{D}_{3h}$ symmetry for the Hoyle state itself.
This is in contrast to some studies which suggest a ``bent-arm'' (obtuse-triangle) structure for the Hoyle state, resembling $\mathrm{^{8}Be} + \alpha$ configurations \cite{PhysRevLett.106.192501,PhysRevLett.98.032501}.
Given the role of the Hoyle state in understanding $\alpha$ clustering, determining if it has a breathing-mode excitation is of great importance.

Identification of this predicted breathing-mode excitation is complicated by phenomenological and experimental factors. 
The ghost of the Hoyle state extends some considerable energy above the main peak of the Hoyle state, overlying the region in which the breathing-mode is predicted and resulting in interference effects with higher-lying monopole resonances \cite{PhysRevC.81.024303}.
The region above the Hoyle state was studied by Itoh \textit{et al.} through the $\mathrm{^{12}C}$($\alpha,\alpha^\prime$)$\mathrm{^{12}C}$ reaction and a peak-fitting analysis with Gaussian lineshapes required an additional peak at $E_{x} \approx 9.04(9)$ MeV with $\Gamma = 1.45(18)$ MeV \cite{PhysRevC.84.054308}.
However, the Gaussian lineshapes employed in Ref. \cite{PhysRevC.84.054308} do not capture the behavior of near-threshold resonances or interference effects.
Since a multipole decomposition analysis (MDA) revealed the region at $E_{x} \approx 9$ MeV to be predominantly monopole, a number of authors \cite{PhysRevC.93.054307, PhysRevC.94.044319, 10.1093/ptep/ptt048, 10.1093/ptep/ptw178} have discussed the additional Gaussian peak in the context of the breathing-mode excitation of the Hoyle state.
A more detailed analysis is required, taking into account the complex shape of the Hoyle state and potential interference effects between resonances.
The objective of this work is to investigate sources of monopole strength in $^{12}$C between $E_{x} = 7$ and $13$ MeV using a consistent analysis of $\mathrm{^{12}C}(\alpha, \alpha^{\prime})\mathrm{^{12}C}$ and $\mathrm{^{14}C}(p, t)\mathrm{^{12}C}$ reaction data. 
The intention is to determine if the two previously established sources of monopole strength and the interference between them can reproduce the experimental data, or if an additional source of monopole strength is required.

To populate the excitation-energy region of interest, measurements of the $\mathrm{^{12}C}(\alpha, \alpha^{\prime})\mathrm{^{12}C}$ and $\mathrm{^{14}C}(p, t)\mathrm{^{12}C}$ reactions at divers laboratory angles and beam energies were performed using the K600 spectrometer at the iThemba Laboratory for Accelerator-Based Sciences (iThemba LABS) in South Africa.
The experimental conditions are summarized in Table \ref{tab:SummaryOfExperimentalParameters} and a comprehensive description is reported in Ref. \cite{li2020multiprobe}.
\begin{table}[b]
\caption{\label{tab:SummaryOfExperimentalParameters}%
Summary of experimental parameters. 
}
\begin{small}
\begin{threeparttable}
\begin{tabular}{lD{.}{.}{2.0}ccc}
\hline\hline
\multicolumn{1}{c}{Reaction} &
\multicolumn{1}{c}{Angle} &
\multicolumn{1}{c}{$E_{\beam}$} &
\multicolumn{1}{c}{Target} &
\multicolumn{1}{c}{Fitted $E_{x}$} \\
&
\multicolumn{1}{c}{[deg]} &
\multicolumn{1}{c}{[MeV]} &
($\mu$g/cm$^{2}$) &
range [MeV]
\vspace{2pt} \\
\hline \\ [-2.3ex]
$\mathrm{^{12}C}(\alpha, \alpha^{\prime})\mathrm{^{12}C}$   & 0  & 118  & \,\,\,$\mathrm{^{nat}C}$ (1053) & 5.0\hphantom{1} - 14.8 \\
                                                            & 0  & 160  & $\mathrm{^{nat}C}$ (300)  & 7.3\hphantom{1} - 20.0 \\
                                                            & 0  & 200  & $\mathrm{^{nat}C}$ (290)   & N.A.\tnote{a} \\
                                                            & 10 & 196  & $\mathrm{^{nat}C}$ (290)  & 7.15 - 21.5 \\
\hline \\ [-2.3ex]
$\mathrm{^{14}C}(p, t)\mathrm{^{12}C}$  & 0     & 100       & $\mathrm{^{14}C}$ (280)   & 6.0\hphantom{1} - 15.3 \\
                                        & 21    & 67.5      & $\mathrm{^{14}C}$ (300)   & 6.8\hphantom{1} - 14.5 \\
\hline\hline
\end{tabular}
    \begin{tablenotes}
      \item[a]{Only the charged-particle decays were analyzed as the Hoyle state was not fully accepted on the focal plane.}
    \end{tablenotes}
\end{threeparttable}
\end{small}
\end{table}
Measurements with different selectivity were chosen to exploit differences in population strength between contributing broad states, and enable a self-consistent, simultaneous analysis of the inclusive spectra.
The $\mathrm{^{12}C}(\alpha, \alpha^{\prime})\mathrm{^{12}C}$ reaction at $\theta_{\textrm{lab}} = 0^{\circ}$ strongly populates collective monopole excitations, in contrast to the $\mathrm{^{14}C}(p, t)\mathrm{^{12}C}$ reaction which is less selective to collective isoscalar monopole excitations \cite{GargU, Harakeh_MN_vanDerWoude_A, PhysRevC.83.037302}.
Proton and $\alpha$-particle beams were extracted from the separated-sector cyclotron and transported down a dispersion-matched beamline to the target position of the K600 magnetic spectrometer \cite{NEVELING201129}.
Ejectiles were momentum-analyzed in the spectrometer and detected at the focal plane in a modular combination of vertical drift chambers and plastic scintillators.
Coincident charged-particle decay from excited $\mathrm{^{12}C}$ states were detected in the \textsc{cake}, an array of double-sided silicon strip detectors \cite{Adsley:2016blb}, for the measurements of $\mathrm{^{12}C}(\alpha, \alpha^{\prime})\mathrm{^{12}C}$ at $\theta_{\textrm{lab}} = 0^{\circ}$ ($E_{\alpha}$ = 200 MeV) and $\mathrm{^{14}C}(p, t)\mathrm{^{12}C}$ $\theta_{\textrm{lab}} = 0^{\circ}$.

\begin{figure}[ht]
\includegraphics[width=\columnwidth]{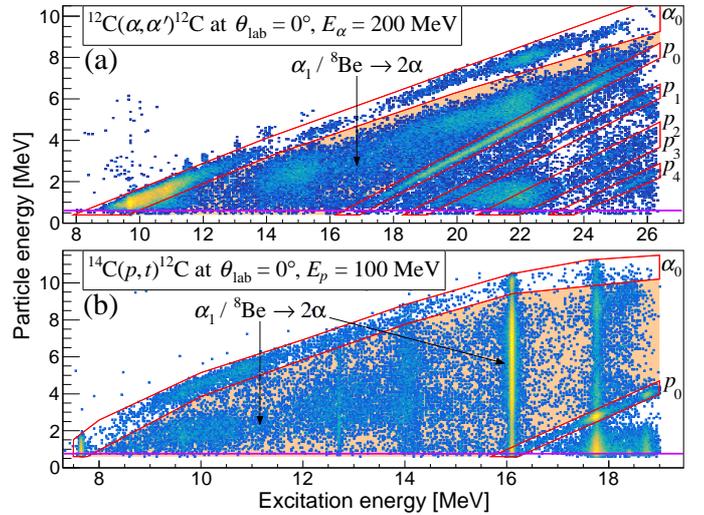}
\caption{\label{fig:ChargedParticleDecay_SiliconEnergyVsExcitationEnergy_letter} 
(Color online) Matrices of decay-particle energy versus the excitation energy of $\mathrm{^{12}C}$.
Red lines indicate the observed $\alpha_{0}$ and proton decay modes of $\mathrm{^{12}C}$.
The excitation energy of the $\mathrm{^{12}C}$ recoil is inferred from the momentum of the $\alpha$/triton ejectile measured at the focal plane.
$\alpha_{0}$ and $\alpha_{1}$ denote $\alpha$ decay to the ground and first-excited states of $\mathrm{^{8}Be}$, respectively.
$p_{0}$ and $p_{1}$ denote proton decay to the ground and first-excited states of $\mathrm{^{11}B}$, respectively (and so forth).
Loci within the orange-highlighted regions correspond to $\alpha$ particles from either the $\alpha_{1}$ decay of $\mathrm{^{12}C}$ or $\mathrm{^{8}Be} \rightarrow 2\alpha$ breakup.
Violet, horizontal lines indicate the approximate electronic thresholds for the \textsc{cake}.
}
\end{figure}

The matrices of decay-particle energy (detected in \textsc{cake}) vs. excitation energy are shown in Fig. \ref{fig:ChargedParticleDecay_SiliconEnergyVsExcitationEnergy_letter}: loci corresponding to the $\alpha$ and proton ($p$) decay modes are observed, with $\alpha_{0}$ and $\alpha_{1}$ denoting $\alpha$ decay to the ground and first-excited states of $\mathrm{^{8}Be}$, respectively.

\begin{figure}[ht!]
\includegraphics[width=\columnwidth]{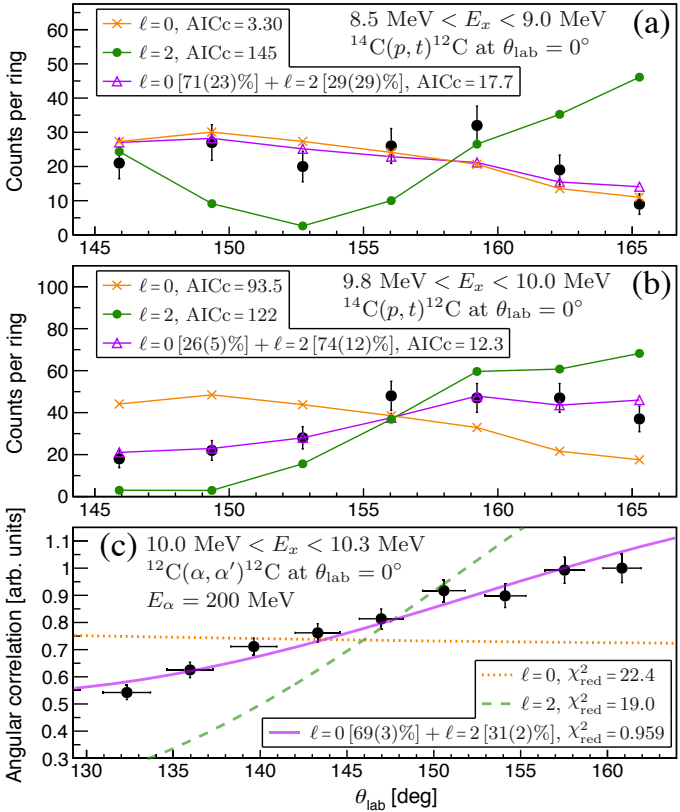}
\caption{\label{fig:AngularDistributions_PR226_PR240_differentOrdering_letter_overpic}
Angular correlations of $\alpha_{0}$ decay obtained in the $\mathrm{^{12}C}(\alpha, \alpha^{\prime})\mathrm{^{12}C}$ and $\mathrm{^{14}C}(p, t)\mathrm{^{12}C}$ reactions at bombarding energies of 100 and 200 MeV, respectively.
For distributions with low counts, the AICc estimator was used to determine the best quality model \cite{Konishi:2007:ICS:1554702, Cavanaugh_1997}.
}
\end{figure}

For $E_{x} = $ 8.5 to 9.0 MeV populated with the $\mathrm{^{14}C}(p, t)\mathrm{^{12}C}$ reaction at $\theta_{\textrm{lab}} = 0^{\circ}$, the $\alpha_{0}$ angular correlation in Fig. \ref{fig:AngularDistributions_PR226_PR240_differentOrdering_letter_overpic}(a) is predominantly described by isotropic decay.
This is consistent with Ref. \cite{PhysRevC.84.054308} indicating that this region is dominated by monopole strength.
For the $\mathrm{^{12}C}(\alpha, \alpha^{\prime})\mathrm{^{12}C^{*}}$ reaction, the $\alpha_{0}$ decay cannot be reliably measured below $E_{x} \approx 10$ MeV since the $\alpha$ particles fall below the threshold of the silicon detectors (see Fig. \ref{fig:ChargedParticleDecay_SiliconEnergyVsExcitationEnergy_letter}). 
For the $\mathrm{^{14}C}(p, t)\mathrm{^{12}C}$ reaction, the momentum boost of the recoiling $\mathrm{^{12}C}$ nucleus means that $\alpha$ particles have a higher laboratory energy and can be detected down to a lower corresponding excitation energy.

Figs. \ref{fig:AngularDistributions_PR226_PR240_differentOrdering_letter_overpic}(b) and \ref{fig:AngularDistributions_PR226_PR240_differentOrdering_letter_overpic}(c) show the angular correlations for $\alpha_{0}$ decay in the regions $E_{x} = $ 9.8 to 10.0 MeV and $E_{x} = $ 10.0 to 10.3 MeV for the $\mathrm{^{14}C}(p, t)\mathrm{^{12}C}$ and $\mathrm{^{12}C}(\alpha, \alpha^{\prime})\mathrm{^{12}C}$ reactions, respectively.
For both cases, the data are well reproduced by an incoherent sum of $\ell=0$ and $\ell=2$ decay components which are of the same order of magnitude.
This result confirms the existence of the $2_{2}^{+}$ state in agreement with the $\mathrm{^{12}C}(\alpha, \alpha^{\prime})\mathrm{^{12}C}$ measurements of Itoh \textit{et al.} \cite{PhysRevC.84.054308} and Freer \textit{et al.} \cite{PhysRevC.86.034320}.
In summary, the angular correlations show that the $E_{x} \approx 9$ MeV region is primarily monopolar in nature, and that a consistent description of the $E_{x} \approx 10$ MeV region must include the $2_{2}^{+}$ state with a strength comparable to the broad monopole contributions at $E_{x} \approx 10$ MeV for the $\mathrm{^{14}C}(p, t)\mathrm{^{12}C}$ and $\mathrm{^{12}C}(\alpha, \alpha^{\prime})\mathrm{^{12}C}$ reactions at $\theta_{\textrm{lab}} = 0^{\circ}$.

\begin{figure}[hb!]
\includegraphics[width=\columnwidth]{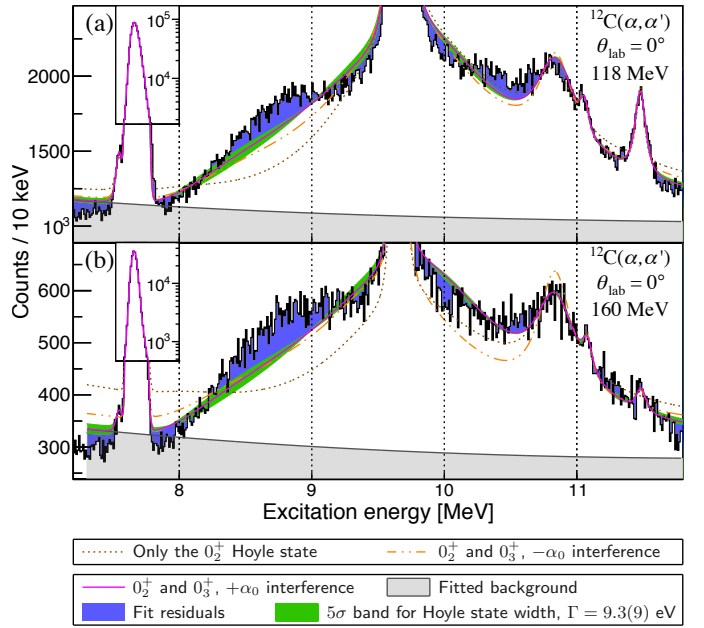}
\caption{ \label{fig:AssembledFittedSpectra_MX_PS1_truncated_letter_overpic}
(Color online) Comparison of fits, using only previously established resonances, for the inclusive inelastically scattered alpha yields of the $\mathrm{^{12}C}(\alpha, \alpha^{\prime})\mathrm{^{12}C}$ reaction at $\theta_{\textrm{lab}} = 0^{\circ}$ with different beam energies.
}
\end{figure}

\begin{figure}[t!]
\includegraphics[width=\columnwidth]{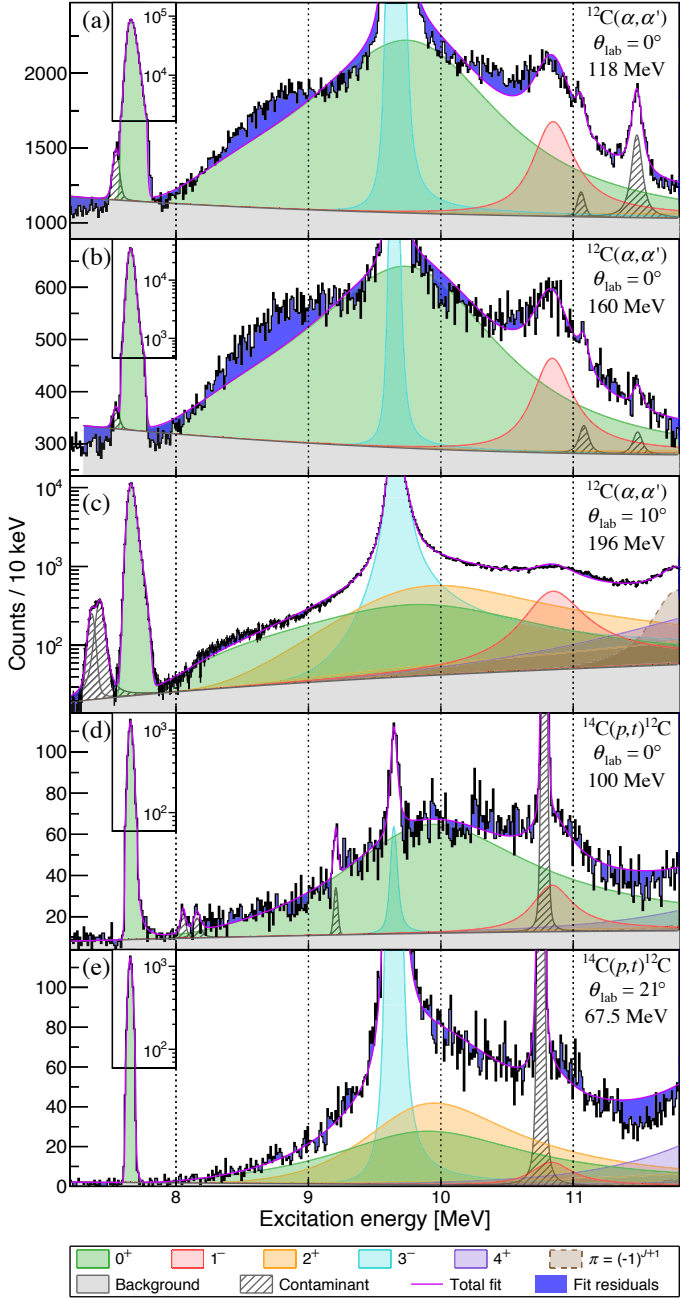}
\caption{ \label{fig:AssembledFittedSpectra_M2_PS1_letter} 
(Color online) Decomposition of the optimal fit which only includes previously established resonances: the $0_{2}^{+}$ and $0_{3}^{+}$ states constructively interfere in the $\alpha_{0}$ channel (see Fig. \ref{fig:AssembledFittedSpectra_MX_PS1_truncated_letter_overpic}).
Panels (a) - (e) correspond to the inclusive ejectile yields for the employed $\mathrm{^{12}C}(\alpha, \alpha^{\prime})\mathrm{^{12}C}$ and $\mathrm{^{14}C}(p, t)\mathrm{^{12}C}$ reactions at various measurement angles and beam energies.
}
\end{figure}

The inclusive excitation-energy spectra from five different measurements (see Table 1) were simultaneously analyzed with phenomenological lineshape parameterizations from multi-level, multi-channel \textbf{R}-matrix theory.
Only the $\alpha_{0}$ and $\alpha_{1}$ decay modes are considered; the proton-decay channel is not open at the excitation energies of interest for this work.
For $\alpha_{1}$ decay, the penetrability accounts for the broad width of the first-excited $2_{1}^{+}$ state of $\mathrm{^{8}Be}$.
For decays with multiple possible angular momenta of decay, the lowest $\ell$-value of decay is assumed to dominate and is set as the exclusive channel.
Instrumental backgrounds were simultaneously fitted. 
Experimental factors, such as the spectrometer ion optics and the response of the drift chambers, were included in the analyses of the spectra \cite{li2020multiprobe}.
Feeding factors capturing the excitation-energy dependence for each incoming reaction channel were determined with \textsc{chuck3} \cite{CHUCK3}.
Fig. \ref{fig:AssembledFittedSpectra_MX_PS1_truncated_letter_overpic} presents the optimized \textbf{R}-matrix fits, which only include the previously established resonances between $E_{x} = 7$ and 13 MeV (as well as contaminant peaks), where the Hoyle-state width has been fixed at $\Gamma = 9.3$ eV, for the measurements of $\mathrm{^{12}C}(\alpha, \alpha^{\prime})\mathrm{^{12}C}$ at $\theta_{\textrm{lab}} = 0^{\circ}$---a reaction that is highly selective for collective monopole strengths.
Including only the Hoyle state provides a poor description of the data, with a large underestimation of the experimental spectra at $E_{x} \approx 9$ MeV.
Similarly, the models including both the $0_{2}^{+}$ and $0_{3}^{+}$ states without interference and with destructive $\alpha_{0}$ interference (denoted $-\alpha_{0}$) underestimate the data at $E_{x} \approx 9$ MeV.
The optimal model with all previously established states is achieved by including both the $0_{2}^{+}$ and $0_{3}^{+}$ monopole resonances with constructive $\alpha_{0}$ interference (denoted $+\alpha_{0}$), however the systematic underestimation of the data at $E_{x} \approx 9$ MeV remains.
A decomposition of the best fit, shown in Fig. \ref{fig:AssembledFittedSpectra_M2_PS1_letter}, indicates a highly suppressed strength for the $2_{2}^{+}$ state located at $E_{x} = 9.870(60)$ MeV.
This is inconsistent with the charged-particle decay data discussed earlier and shown in Fig. \ref{fig:AngularDistributions_PR226_PR240_differentOrdering_letter_overpic}, which indicate significant $2_{2}^{+}$ strength at $E_{x} \approx 10$ MeV, as well as the analyses of $\mathrm{^{12}C}(\alpha, \alpha^{\prime})\mathrm{^{12}C}$ presented in Refs. \cite{PhysRevC.84.054308, PhysRevC.86.034320}.
To test whether a difference in the total width of the Hoyle state may improve the fit, the total width of the Hoyle state was tested at $5\sigma$ below and above the listed value of $\Gamma = 9.3(9)$ eV \cite{ENSDF}, with the $5\sigma$ band in Fig. \ref{fig:AssembledFittedSpectra_MX_PS1_truncated_letter_overpic} corresponding to the range spanned by the associated fits.
Even at these extreme values, a clear systematic excess in the data remains in the dominantly monopole region at $E_{x} \approx 9$ MeV.

\begin{figure}[t!]
\includegraphics[width=\columnwidth]{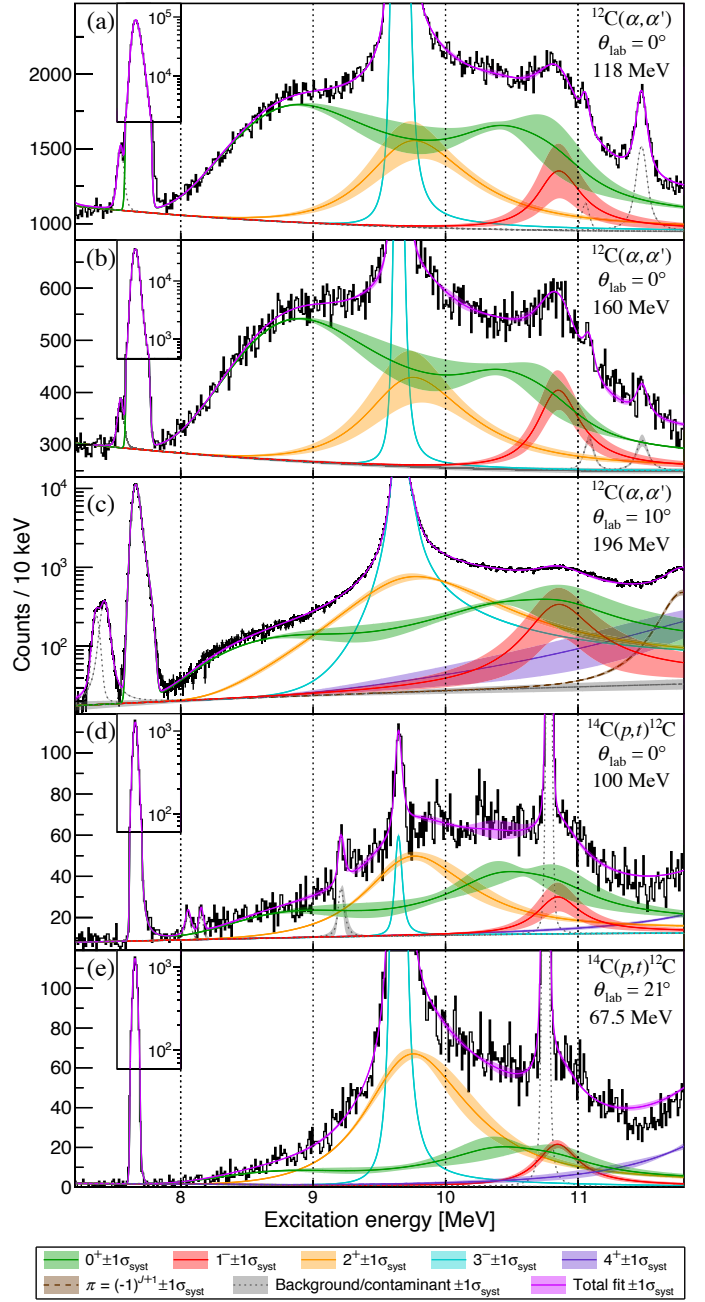}
\caption{ \label{fig:AssembledFittedSpectra_M3_PS1_3LA_letter} 
(Color online) Decomposition of the optimal fit which includes the previously established $0_{2}^{+}$ and $0_{3}^{+}$ states as well as the additional $0_{\!\Delta}^{+}$ resonance at $E_{x} \approx 9$ MeV.
Panels (a) - (e) correspond to the inclusive ejectile yields for the employed $\mathrm{^{12}C}(\alpha, \alpha^{\prime})\mathrm{^{12}C}$ and $\mathrm{^{14}C}(p, t)\mathrm{^{12}C}$ reactions at various measurement angles and beam energies.
} 
\end{figure}

To account for the excess strength at $E_{x} \approx 9$ MeV, which according to both the charged-particle decay data of this work and Ref. \cite{PhysRevC.84.054308} is monopolar in nature, an additional monopole state was introduced, denoted $0_{\!\Delta}^{+}$.
Two parameterizations for the resulting monopole strength were investigated: in the first, the $0_{\!\Delta}^{+}$ state was permitted to interfere with the previously established $0_{2}^{+}$ and $0_{3}^{+}$ states.
In the second, interference involving the $0_{\!\Delta}^{+}$ state was omitted whilst the $0_{2}^{+}$ and $0_{3}^{+}$ states were still permitted to interfere.
The optimized fit for the first parameterization is presented in Fig. \ref{fig:AssembledFittedSpectra_M3_PS1_3LA_letter}, which produces a significantly improved description of the data, particularly at $E_{x} \approx 9$ MeV.
The $1\sigma$ bands on Fig. \ref{fig:AssembledFittedSpectra_M3_PS1_3LA_letter} show the systematic dispersion of lineshapes corresponding to different permutations of interference and channel-radii.
The decomposition of the fit reveals contributions from the $2_{2}^{+}$ state which are comparable to the broad monopole strength at $E_{x} \approx 10$ MeV for the measurements of $\mathrm{^{12}C}(\alpha, \alpha^{\prime})\mathrm{^{12}C}$ and $\mathrm{^{14}C}(p, t)\mathrm{^{12}C}$ at $\theta_{\textrm{lab}} = 0^{\circ}$---a result consistent with the present charged-particle decay data and Ref. \cite{PhysRevC.84.054308}.
Furthermore, the fit reveals the overall monopole strength to be double peaked in the region $E_{x} = 9-11$ MeV for the $\mathrm{^{12}C}(\alpha, \alpha^{\prime})\mathrm{^{12}C}$ reaction ($\theta_{\textrm{lab}} = 0^{\circ}$), in agreement with Ref. \cite{PhysRevC.84.054308}.
The additional $0_{\!\Delta}^{+}$ state has best fit parameters of $E_{r} = 9.566\,\mypm\,0.018_{\textrm{(stat)}}\,\mypm\,0.104_{\textrm{(syst)}}$ MeV with $\Gamma(E_{r}) = 3.203\,\mypm\,0.061_{\textrm{(stat)}}\,\mypm\,0.599_{\textrm{(syst)}}$ MeV.
The second parameterization, which omitted possible interference involving the $0_{\!\Delta}^{+}$ state, yielded $E_{r} = 9.379\,\mypm\,0.013_{\textrm{(stat)}}\,\mypm\,0.050_{\textrm{(syst)}}$ MeV with $\Gamma(E_{r}) = 4.565\,\mypm\,0.022_{\textrm{(stat)}}\,\mypm\,0.107_{\textrm{(syst)}}$ MeV with $\Gamma_{\FWHM} = 2.066\,\mypm\,0.005_{\textrm{(stat)}}\,\mypm\,0.098_{\textrm{(syst)}}$ MeV \cite{li2020multiprobe}.

There is no consensus on the number and properties of the monopole states in this astrophysically important excitation-energy region. 
This is due to the variety of analysis methods used to describe the data, some of which do not include the strong threshold effects or interference. 
However, there is a rough similarity between the additional $0_{\!\Delta}^{+}$ state in this work to a structure in Ref. \cite{PhysRevC.84.054308} that was described with a Gaussian peak.
Since the peak-fitting analysis with Gaussian lineshapes in Ref. \cite{PhysRevC.84.054308} did not account for the physical effects of near-threshold resonances (such as the Hoyle state's ghost) or interference, care must be taken in this comparison since the two analyses are based on different foundations, resulting in different relative population strengths.
Good agreement with recent measurements \cite{PhysRevC.81.024303,RevModPhys.90.035004} of the $2_{2}^{+}$ and $0_{3}^{+}$ states was also achieved in this work (See Ref. \cite{li2020multiprobe} for details).

From the analysis above, there is clear evidence for an additional source of monopole strength in the excitation-energy region of $E_{x} = 7$ to 13 MeV; the data cannot be reproduced with only the previously established $0_{2}^{+}$ and $0_{3}^{+}$ states within the acceptable model space for the parameters of those states.
The introduction of a new monopole state significantly improved the description of the data, including calculated relative strengths for the new monopole state and the $2_{2}^{+}$ state in agreement with the angular correlations from charged-particle decay in the present work and the multipole decomposition analysis of Ref. \cite{PhysRevC.84.054308}.
The parameters of this new monopole state are in good agreement with theoretical predictions \cite{PhysRevC.71.021301, CKurokawa, 10.1093/ptep/ptt048, PhysRevC.94.044319, PhysRevC.99.064327}, making this state the leading candidate for the breathing-mode excitation of the Hoyle state.

That a significant excess of monopole strength is consistently observed at $E_{x} \approx 9$ MeV with $\mathrm{^{12}C}(\alpha, \alpha^{\prime})\mathrm{^{12}C}$ at $\theta_{\textrm{lab}} = 0^{\circ}$, but not with $\mathrm{^{14}C}(p, t)\mathrm{^{12}C}$ at the same angle is suggestive of a collective character for this additional monopole state. 
The former reaction is considerably more selective for collective isoscalar monopole excitations than the latter \cite{GargU, Harakeh_MN_vanDerWoude_A, PhysRevC.83.037302}.
The interpretation of the additional $0_{\!\Delta}^{+}$ state as a breathing-mode excitation is consistent with its weak population in beta-decay data \cite{PhysRevC.81.024303}, as $\beta$-decay is less favourable for transitions involving radial changes in comparison to other nuclear/electromagnetic probes \cite{PhysRevLett.116.132501, PhysRevC.86.024312}.
More evidence that the newly observed state should be identified as the breathing-mode excitation of the Hoyle state comes from the relative ordering of resonance energies, with the predicted excitation energy around 2 MeV above the Hoyle state and below the higher-lying ($E_{x} \gtrapprox 10$ MeV) broad monopole strengths which have been interpreted to correspond to `bent-arm'' ($\mathrm{^{8}Be} + \alpha$) configurations \cite{PhysRevLett.106.192501,PhysRevLett.98.032501} and/or large-amplitude breathing-mode excitations \cite{PhysRevC.82.034307}.


It is possible that the excess monopole strength at $E_{x} \approx 9$ MeV can be alternatively explained by the artifacts of the current phenomenological descriptions of how the Hoyle-state properties evolve with excitation energy.
Phenomenological \textbf{R}-matrix models employed in astrophysics typically assume that only two-body effects are important \cite{HOUFynbo_2005,RevModPhys.30.257}.
Whilst the direct $3\alpha$ decay branch has been shown to be small at the primary peak of the Hoyle state \cite{PhysRevLett.119.132501, PhysRevLett.119.132502, TKRana_2019, PhysRevC.102.041303}, at higher excitation energies, the direct branch may be non-negligible due to the enhanced penetrability through the Coulomb barrier (see Ref. \cite{PhysRevC.101.021302}).
More sophisticated parameterizations of the monopole strength \cite{RSmith_2019} should be explored. 
Inaccurate parameterizations may result in inaccurate extraction of the properties of states in $\mathrm{^{12}C}$, affecting the $3\alpha$ reaction rate particularly above 2 GK in, for example, the shock front for type II supernovae \cite{HOUFynbo_2005, Pruet_2005, Frohlich_2006}.
In such environments, improved theoretical descriptions, including the additional $0_{\!\Delta}^{+}$ state, may modify the triple-$\alpha$ reaction rate.
A quantitative estimation is beyond the scope of this work as knowledge on the radiative widths for the broad contributions above the Hoyle state is required but is not available from these data.
The inclusive spectra studied in this work are most sensitive to the resonance energies and total widths of resonances, the latter of which are dominated by charged-particle decay channels in the excitation-energy region of interest between $E_{x} = 7$ and $13$ MeV.
The radiative width for the rotational $2_{2}^{+}$ excitation of the Hoyle state --- a crucial quantity for the triple-$\alpha$ process at ${T_{9} \sim 4}$ --- has been measured \cite{PhysRevLett.110.152502}. 
This state is not of a similar nature to the collective $0_{\!\Delta}^{+}$ state, therefore no useful information on the reduced transition probabilities can be extracted from this comparison.

In summary, despite the above discussed aspects requiring further investigation, a leading candidate for the breathing mode excitation of the Hoyle state has been observed in additional monopole strength in $^{12}$C at $E_\mathrm{x} \approx 9$ MeV. 
The evidence for the breathing mode at this excitation energy favors the Hoyle state being an equilateral triangle.

\section*{Acknowledgements}

This work is based on the research supported in part by the National Research Foundation of South Africa (Grant Numbers: 85509, 86052, 118846, 90741).
The authors acknowledge the accelerator staff of iThemba LABS for providing excellent beams.
The authors would like to thank E.~Khan, A.~Tamii, B.~Zhou, K.~Masaaki, H.O.U.~Fynbo, M.~Itoh and J.~Carter for useful discussions.
PA acknowledges support from the Claude Leon Foundation in the form of a postdoctoral fellowship.
The computations were performed on resources provided by UNINETT Sigma2 - the National Infrastructure for High Performance Computing and Data Storage in Norway.
The authors are grateful to A.C. Larsen, F. Zeiser and F. Pogliano for their assistance with UNINETT Sigma2.










\bibliography{LowLyingMonopoleStrengthsIn12C}

\begin{thebibliography}{10}
\expandafter\ifx\csname url\endcsname\relax
  \def\url#1{\texttt{#1}}\fi
\expandafter\ifx\csname urlprefix\endcsname\relax\def\urlprefix{URL }\fi
\expandafter\ifx\csname href\endcsname\relax
  \def\href#1#2{#2} \def\path#1{#1}\fi

\bibitem{BoseSN}
S.~N. Bose, \href{https://doi.org/10.1007/BF01327326}{Plancks gesetz und
  lichtquantenhypothese}, Zeitschrift f{\"u}r Physik 26~(1) (1924) 178--181.
\newblock \href {http://dx.doi.org/10.1007/BF01327326}
  {\path{doi:10.1007/BF01327326}}.
\newline\urlprefix\url{https://doi.org/10.1007/BF01327326}

\bibitem{EinsteinA}
A.~Einstein, Quantentheorie des einatomigen idealen gases, Sitzungsberichte der
  Preussischen Akademie der Wissenschaften 1 (1925) 3.

\bibitem{FREER20141}
M.~Freer, H.~Fynbo,
  \href{http://www.sciencedirect.com/science/article/pii/S0146641014000453}{{The
  {H}oyle state in $^{12}$C}}, Progress in Particle and Nuclear Physics 78
  (2014) 1 -- 23.
\newblock \href {http://dx.doi.org/https://doi.org/10.1016/j.ppnp.2014.06.001}
  {\path{doi:https://doi.org/10.1016/j.ppnp.2014.06.001}}.
\newline\urlprefix\url{http://www.sciencedirect.com/science/article/pii/S0146641014000453}

\bibitem{PhysRevLett.113.102501}
M.~Itoh, S.~Ando, T.~Aoki, H.~Arikawa, S.~Ezure, K.~Harada, T.~Hayamizu,
  T.~Inoue, T.~Ishikawa, K.~Kato, H.~Kawamura, Y.~Sakemi, A.~Uchiyama,
  \href{https://link.aps.org/doi/10.1103/PhysRevLett.113.102501}{Further
  improvement of the upper limit on the direct $3\ensuremath{\alpha}$ decay
  from the {H}oyle state in $^{12}\mathrm{C}$}, Phys. Rev. Lett. 113 (2014)
  102501.
\newblock \href {http://dx.doi.org/10.1103/PhysRevLett.113.102501}
  {\path{doi:10.1103/PhysRevLett.113.102501}}.
\newline\urlprefix\url{https://link.aps.org/doi/10.1103/PhysRevLett.113.102501}

\bibitem{PhysRevLett.119.132501}
D.~Dell'Aquila, I.~Lombardo, G.~Verde, M.~Vigilante, L.~Acosta, C.~Agodi,
  F.~Cappuzzello, D.~Carbone, M.~Cavallaro, S.~Cherubini, A.~Cvetinovic,
  G.~D'Agata, L.~Francalanza, G.~L. Guardo, M.~Gulino, I.~Indelicato,
  M.~La~Cognata, L.~Lamia, A.~Ordine, R.~G. Pizzone, S.~M.~R. Puglia, G.~G.
  Rapisarda, S.~Romano, G.~Santagati, R.~Spart\`a, G.~Spadaccini, C.~Spitaleri,
  A.~Tumino,
  \href{https://link.aps.org/doi/10.1103/PhysRevLett.119.132501}{High-precision
  probe of the fully sequential decay width of the {H}oyle state in
  $^{12}\mathrm{C}$}, Phys. Rev. Lett. 119 (2017) 132501.
\newblock \href {http://dx.doi.org/10.1103/PhysRevLett.119.132501}
  {\path{doi:10.1103/PhysRevLett.119.132501}}.
\newline\urlprefix\url{https://link.aps.org/doi/10.1103/PhysRevLett.119.132501}

\bibitem{PhysRevLett.119.132502}
R.~Smith, T.~Kokalova, C.~Wheldon, J.~E. Bishop, M.~Freer, N.~Curtis, D.~J.
  Parker, \href{https://link.aps.org/doi/10.1103/PhysRevLett.119.132502}{New
  measurement of the direct $3\ensuremath{\alpha}$ decay from the
  $^{12}\mathrm{C}$ {H}oyle state}, Phys. Rev. Lett. 119 (2017) 132502.
\newblock \href {http://dx.doi.org/10.1103/PhysRevLett.119.132502}
  {\path{doi:10.1103/PhysRevLett.119.132502}}.
\newline\urlprefix\url{https://link.aps.org/doi/10.1103/PhysRevLett.119.132502}

\bibitem{TKRana_2019}
T.~K. Rana, S.~Bhattacharya, C.~Bhattacharya, S.~Manna, S.~Kundu, K.~Banerjee,
  R.~Pandey, P.~Roy, A.~Dhal, G.~Mukherjee, V.~Srivastava, A.~Dey,
  A.~Chaudhuri, T.~K. Ghosh, A.~Sen, M.~A. Asgar, T.~Roy, J.~K. Sahoo, J.~K.
  Meena, A.~K. Saha, R.~M. Saha, M.~Sinha, A.~Roy,
  \href{http://www.sciencedirect.com/science/article/pii/S0370269319302618}{New
  high precision study on the decay width of the {H}oyle state in
  $^{12}\mathrm{C}$}, Physics Letters B 793 (2019) 130--133.
\newblock \href
  {http://dx.doi.org/https://doi.org/10.1016/j.physletb.2019.04.028}
  {\path{doi:https://doi.org/10.1016/j.physletb.2019.04.028}}.
\newline\urlprefix\url{http://www.sciencedirect.com/science/article/pii/S0370269319302618}

\bibitem{PhysRevC.101.021302}
R.~Smith, M.~Gai, M.~W. Ahmed, M.~Freer, H.~O.~U. Fynbo, D.~Schweitzer, S.~R.
  Stern, \href{https://link.aps.org/doi/10.1103/PhysRevC.101.021302}{Stringent
  upper limit on the direct $3\ensuremath{\alpha}$ decay of the {H}oyle state
  in $^{12}\mathrm{C}$}, Phys. Rev. C 101 (2020) 021302.
\newblock \href {http://dx.doi.org/10.1103/PhysRevC.101.021302}
  {\path{doi:10.1103/PhysRevC.101.021302}}.
\newline\urlprefix\url{https://link.aps.org/doi/10.1103/PhysRevC.101.021302}

\bibitem{PhysRevC.102.041303}
J.~Bishop, G.~V. Rogachev, S.~Ahn, E.~Aboud, M.~Barbui, A.~Bosh, C.~Hunt,
  H.~Jayatissa, E.~Koshchiy, R.~Malecek, S.~T. Marley, E.~C. Pollacco, C.~D.
  Pruitt, B.~T. Roeder, A.~Saastamoinen, L.~G. Sobotka, S.~Upadhyayula,
  \href{https://link.aps.org/doi/10.1103/PhysRevC.102.041303}{Almost
  medium-free measurement of the hoyle state direct-decay component with a
  tpc}, Phys. Rev. C 102 (2020) 041303.
\newblock \href {http://dx.doi.org/10.1103/PhysRevC.102.041303}
  {\path{doi:10.1103/PhysRevC.102.041303}}.
\newline\urlprefix\url{https://link.aps.org/doi/10.1103/PhysRevC.102.041303}

\bibitem{PhysRevLett.125.182701}
T.~Kib\'edi, B.~Alshahrani, A.~E. Stuchbery, A.~C. Larsen, A.~G\"orgen,
  S.~Siem, M.~Guttormsen, F.~Giacoppo, A.~I. Morales, E.~Sahin, G.~M. Tveten,
  F.~L.~B. Garrote, L.~C. Campo, T.~K. Eriksen, M.~Klintefjord, S.~Maharramova,
  H.-T. Nyhus, T.~G. Tornyi, T.~Renstr\o{}m, W.~Paulsen,
  \href{https://link.aps.org/doi/10.1103/PhysRevLett.125.182701}{Radiative
  width of the hoyle state from $\ensuremath{\gamma}$-ray spectroscopy}, Phys.
  Rev. Lett. 125 (2020) 182701.
\newblock \href {http://dx.doi.org/10.1103/PhysRevLett.125.182701}
  {\path{doi:10.1103/PhysRevLett.125.182701}}.
\newline\urlprefix\url{https://link.aps.org/doi/10.1103/PhysRevLett.125.182701}

\bibitem{PhysRevC.102.024320}
T.~K. Eriksen, T.~Kib\'edi, M.~W. Reed, A.~E. Stuchbery, K.~J. Cook, A.~Akber,
  B.~Alshahrani, A.~A. Avaa, K.~Banerjee, A.~C. Berriman, L.~T. Bezzina,
  L.~Bignell, J.~Buete, I.~P. Carter, B.~J. Coombes, J.~T.~H. Dowie,
  M.~Dasgupta, L.~J. Evitts, A.~B. Garnsworthy, M.~S.~M. Gerathy, T.~J. Gray,
  D.~J. Hinde, T.~H. Hoang, S.~S. Hota, E.~Ideguchi, P.~Jones, G.~J. Lane,
  B.~P. McCormick, A.~J. Mitchell, N.~Palalani, T.~Palazzo, M.~Ripper, E.~C.
  Simpson, J.~Smallcombe, B.~M.~A. Swinton-Bland, T.~Tanaka, T.~G. Tornyi,
  M.~O. de~Vries,
  \href{https://link.aps.org/doi/10.1103/PhysRevC.102.024320}{Improved
  precision on the experimental $e0$ decay branching ratio of the hoyle state},
  Phys. Rev. C 102 (2020) 024320.
\newblock \href {http://dx.doi.org/10.1103/PhysRevC.102.024320}
  {\path{doi:10.1103/PhysRevC.102.024320}}.
\newline\urlprefix\url{https://link.aps.org/doi/10.1103/PhysRevC.102.024320}

\bibitem{HOUFynbo_2005}
H.~O.~U. Fynbo, C.~{\relax Aa}. Diget, U.~C. Bergmann, M.~J.~G. Borge,
  J.~Cederk{\"a}ll, P.~Dendooven, L.~M. Fraile, S.~Franchoo, V.~N. Fedosseev,
  B.~R. Fulton, W.~Huang, J.~Huikari, H.~B. Jeppesen, A.~S. Jokinen, P.~Jones,
  B.~Jonson, U.~K{\"o}ster, K.~Langanke, M.~Meister, T.~Nilsson, G.~Nyman,
  Y.~Prezado, K.~Riisager, S.~Rinta-Antila, O.~Tengblad, M.~Turrion, Y.~Wang,
  L.~Weissman, K.~Wilhelmsen, J.~{\"A}yst{\"o}, {The ISOLDE Collaboration},
  \href{https://doi.org/10.1038/nature03219}{Revised rates for the stellar
  triple-$\alpha$ process from measurement of $^{12}\mathrm{C}$ nuclear
  resonances}, Nature 433~(7022) (2005) 136--139.
\newblock \href {http://dx.doi.org/10.1038/nature03219}
  {\path{doi:10.1038/nature03219}}.
\newline\urlprefix\url{https://doi.org/10.1038/nature03219}

\bibitem{PhysRevC.84.054308}
M.~Itoh, H.~Akimune, M.~Fujiwara, U.~Garg, N.~Hashimoto, T.~Kawabata,
  K.~Kawase, S.~Kishi, T.~Murakami, K.~Nakanishi, Y.~Nakatsugawa, B.~K. Nayak,
  S.~Okumura, H.~Sakaguchi, H.~Takeda, S.~Terashima, M.~Uchida, Y.~Yasuda,
  M.~Yosoi, J.~Zenihiro,
  \href{https://link.aps.org/doi/10.1103/PhysRevC.84.054308}{Candidate for the
  2${}^{+}$ excited {H}oyle state at ${E}_{x}\ensuremath{\sim}10$ {M}ev in
  ${}^{12}${C}}, Phys. Rev. C 84 (2011) 054308.
\newblock \href {http://dx.doi.org/10.1103/PhysRevC.84.054308}
  {\path{doi:10.1103/PhysRevC.84.054308}}.
\newline\urlprefix\url{https://link.aps.org/doi/10.1103/PhysRevC.84.054308}

\bibitem{PhysRevC.86.034320}
M.~Freer, M.~Itoh, T.~Kawabata, H.~Fujita, H.~Akimune, Z.~Buthelezi, J.~Carter,
  R.~W. Fearick, S.~V. F\"ortsch, M.~Fujiwara, U.~Garg, N.~Hashimoto,
  K.~Kawase, S.~Kishi, T.~Murakami, K.~Nakanishi, Y.~Nakatsugawa, B.~K. Nayak,
  R.~Neveling, S.~Okumura, S.~M. Perez, P.~Papka, H.~Sakaguchi, Y.~Sasamoto,
  F.~D. Smit, J.~A. Swartz, H.~Takeda, S.~Terashima, M.~Uchida, I.~Usman,
  Y.~Yasuda, M.~Yosoi, J.~Zenihiro,
  \href{https://link.aps.org/doi/10.1103/PhysRevC.86.034320}{{Consistent
  analysis of the 2${}^{+}$ excitation of the $^{12}$C {H}oyle state populated
  in proton and \ensuremath{\alpha}-particle inelastic scattering}}, Phys. Rev.
  C 86 (2012) 034320.
\newblock \href {http://dx.doi.org/10.1103/PhysRevC.86.034320}
  {\path{doi:10.1103/PhysRevC.86.034320}}.
\newline\urlprefix\url{https://link.aps.org/doi/10.1103/PhysRevC.86.034320}

\bibitem{PhysRevLett.110.152502}
W.~R. Zimmerman, M.~W. Ahmed, B.~Bromberger, S.~C. Stave, A.~Breskin,
  V.~Dangendorf, T.~Delbar, M.~Gai, S.~S. Henshaw, J.~M. Mueller, C.~Sun,
  K.~Tittelmeier, H.~R. Weller, Y.~K. Wu,
  \href{https://link.aps.org/doi/10.1103/PhysRevLett.110.152502}{Unambiguous
  identification of the second ${2}^{\mathbf{+}}$ state in $^{12}\mathbf{C}$
  and the structure of the {H}oyle state}, Phys. Rev. Lett. 110 (2013) 152502.
\newblock \href {http://dx.doi.org/10.1103/PhysRevLett.110.152502}
  {\path{doi:10.1103/PhysRevLett.110.152502}}.
\newline\urlprefix\url{https://link.aps.org/doi/10.1103/PhysRevLett.110.152502}

\bibitem{1976AuJPh_29_245B}
F.~C. {Barker}, G.~M. {Crawley}, P.~S. {Miller}, W.~F. {Steele}, {The ghost
  anomaly in the $^{9}$Be(p, d)$^{8}$Be reaction}, Australian Journal of
  Physics 29 (1976) 245.
\newblock \href {http://dx.doi.org/10.1071/PH760245}
  {\path{doi:10.1071/PH760245}}.

\bibitem{PhysRevC.71.021301}
C.~Kurokawa, K.~Kat\ifmmode~\bar{o}\else \={o}\fi{},
  \href{https://link.aps.org/doi/10.1103/PhysRevC.71.021301}{New broad
  ${0}^{+}$ state in $^{12}\mathrm{C}$}, Phys. Rev. C 71 (2005) 021301.
\newblock \href {http://dx.doi.org/10.1103/PhysRevC.71.021301}
  {\path{doi:10.1103/PhysRevC.71.021301}}.
\newline\urlprefix\url{https://link.aps.org/doi/10.1103/PhysRevC.71.021301}

\bibitem{CKurokawa}
C.~Kurokawa, K.~Kat{\=o},
  \href{http://www.sciencedirect.com/science/article/pii/S0375947407005246}{Spectroscopy
  of {$^{12}$C} within the boundary condition for three-body resonant states},
  Nuclear Physics A 792~(1) (2007) 87--101.
\newblock \href
  {http://dx.doi.org/https://doi.org/10.1016/j.nuclphysa.2007.05.007}
  {\path{doi:https://doi.org/10.1016/j.nuclphysa.2007.05.007}}.
\newline\urlprefix\url{http://www.sciencedirect.com/science/article/pii/S0375947407005246}

\bibitem{10.1093/ptep/ptt048}
S.-I. Ohtsubo, Y.~Fukushima, M.~Kamimura, E.~Hiyama,
  \href{https://doi.org/10.1093/ptep/ptt048}{{Complex-scaling calculation of
  three-body resonances using complex-range Gaussian basis functions:
  Application to 3$\alpha$ resonances in $^{12}$C}}, Progress of Theoretical
  and Experimental Physics 2013~(7), 073D02.
\newblock \href
  {http://arxiv.org/abs/http://oup.prod.sis.lan/ptep/article-pdf/2013/7/073D02/19300448/ptt048.pdf}
  {\path{arXiv:http://oup.prod.sis.lan/ptep/article-pdf/2013/7/073D02/19300448/ptt048.pdf}},
  \href {http://dx.doi.org/10.1093/ptep/ptt048}
  {\path{doi:10.1093/ptep/ptt048}}.
\newline\urlprefix\url{https://doi.org/10.1093/ptep/ptt048}

\bibitem{PhysRevC.82.034307}
T.~Furuta, K.~H.~O. Hasnaoui, F.~Gulminelli, C.~Leclercq, A.~Ono,
  \href{https://link.aps.org/doi/10.1103/PhysRevC.82.034307}{Monopole
  oscillations in light nuclei with a molecular dynamics approach}, Phys. Rev.
  C 82 (2010) 034307.
\newblock \href {http://dx.doi.org/10.1103/PhysRevC.82.034307}
  {\path{doi:10.1103/PhysRevC.82.034307}}.
\newline\urlprefix\url{https://link.aps.org/doi/10.1103/PhysRevC.82.034307}

\bibitem{PhysRevC.94.044319}
B.~Zhou, A.~Tohsaki, H.~Horiuchi, Z.~Ren,
  \href{https://link.aps.org/doi/10.1103/PhysRevC.94.044319}{Breathing-like
  excited state of the {H}oyle state in $^{12}\mathrm{C}$}, Phys. Rev. C 94
  (2016) 044319.
\newblock \href {http://dx.doi.org/10.1103/PhysRevC.94.044319}
  {\path{doi:10.1103/PhysRevC.94.044319}}.
\newline\urlprefix\url{https://link.aps.org/doi/10.1103/PhysRevC.94.044319}

\bibitem{PhysRevC.99.064327}
R.~Imai, T.~Tada, M.~Kimura,
  \href{https://link.aps.org/doi/10.1103/PhysRevC.99.064327}{Real-time
  evolution method and its application to the $3\ensuremath{\alpha}$ cluster
  system}, Phys. Rev. C 99 (2019) 064327.
\newblock \href {http://dx.doi.org/10.1103/PhysRevC.99.064327}
  {\path{doi:10.1103/PhysRevC.99.064327}}.
\newline\urlprefix\url{https://link.aps.org/doi/10.1103/PhysRevC.99.064327}

\bibitem{HYLDEGAARD2009459}
S.~Hyldegaard, C.~Forss{\'e}n, C.~{\relax Aa}. Diget, M.~Alcorta, F.~C. Barker,
  B.~Bastin, M.~J.~G. Borge, R.~Boutami, S.~Brandenburg, J.~B{\"u}scher,
  P.~Dendooven, P.~Van~Duppen, T.~Eronen, S.~Fox, B.~R. Fulton, H.~O.~U. Fynbo,
  J.~Huikari, M.~Huyse, H.~B. Jeppesen, A.~Jokinen, B.~Jonson, K.~Jungmann,
  A.~Kankainen, O.~Kirsebom, M.~Madurga, I.~Moore, P.~Navr{\'a}til, T.~Nilsson,
  G.~Nyman, G.~J.~G. Onderwater, H.~Penttil{\"a}, K.~Per{\"a}j{\"a}rvi,
  R.~Raabe, K.~Riisager, S.~Rinta-Antila, A.~Rogachevskiy, A.~Saastamoinen,
  M.~Sohani, O.~Tengblad, E.~Traykov, J.~P. Vary, Y.~Wang, K.~Wilhelmsen, H.~W.
  Wilschut, J.~{\"A}yst{\"o},
  \href{http://www.sciencedirect.com/science/article/pii/S0370269309007771}{Precise
  branching ratios to unbound $^{12}\mathrm{C}$ states from $^{12}\mathrm{N}$
  and $^{12}\mathrm{B}$ $\beta$-decays}, Physics Letters B 678~(5) (2009) 459
  -- 464.
\newblock \href
  {http://dx.doi.org/https://doi.org/10.1016/j.physletb.2009.06.064}
  {\path{doi:https://doi.org/10.1016/j.physletb.2009.06.064}}.
\newline\urlprefix\url{http://www.sciencedirect.com/science/article/pii/S0370269309007771}

\bibitem{PhysRevC.80.044304}
S.~Hyldegaard, C.~{\relax Aa}. Diget, M.~J.~G. Borge, R.~Boutami, P.~Dendooven,
  T.~Eronen, S.~P. Fox, L.~M. Fraile, B.~R. Fulton, H.~O.~U. Fynbo, J.~Huikari,
  H.~B. Jeppesen, A.~S. Jokinen, B.~Jonson, A.~Kankainen, I.~Moore, G.~Nyman,
  H.~Penttil\"a, K.~Per\"aj\"arvi, K.~Riisager, S.~Rinta-Antila, O.~Tengblad,
  Y.~Wang, K.~Wilhelmsen, J.~\"Ayst\"o,
  \href{https://link.aps.org/doi/10.1103/PhysRevC.80.044304}{Branching ratios
  in the $\ensuremath{\beta}$ decays of $^{12}\mathrm{N}$ and
  $^{12}\mathrm{B}$}, Phys. Rev. C 80 (2009) 044304.
\newblock \href {http://dx.doi.org/10.1103/PhysRevC.80.044304}
  {\path{doi:10.1103/PhysRevC.80.044304}}.
\newline\urlprefix\url{https://link.aps.org/doi/10.1103/PhysRevC.80.044304}

\bibitem{ENSDF}
Evaluated Nuclear Structure Data File, http://www.nndc.bnl.gov/ensdf/.

\bibitem{PhysRevLett.106.192501}
E.~Epelbaum, H.~Krebs, D.~Lee, U.-G. Mei\ss{}ner,
  \href{https://link.aps.org/doi/10.1103/PhysRevLett.106.192501}{Ab initio
  calculation of the {H}oyle state}, Phys. Rev. Lett. 106 (2011) 192501.
\newblock \href {http://dx.doi.org/10.1103/PhysRevLett.106.192501}
  {\path{doi:10.1103/PhysRevLett.106.192501}}.
\newline\urlprefix\url{https://link.aps.org/doi/10.1103/PhysRevLett.106.192501}

\bibitem{PhysRevLett.109.252501}
E.~Epelbaum, H.~Krebs, T.~A. L\"ahde, D.~Lee, U.-G. Mei\ss{}ner,
  \href{https://link.aps.org/doi/10.1103/PhysRevLett.109.252501}{Structure and
  rotations of the {H}oyle state}, Phys. Rev. Lett. 109 (2012) 252501.
\newblock \href {http://dx.doi.org/10.1103/PhysRevLett.109.252501}
  {\path{doi:10.1103/PhysRevLett.109.252501}}.
\newline\urlprefix\url{https://link.aps.org/doi/10.1103/PhysRevLett.109.252501}

\bibitem{PhysRevC.61.067305}
R.~Bijker, F.~Iachello,
  \href{https://link.aps.org/doi/10.1103/PhysRevC.61.067305}{Cluster states in
  nuclei as representations of a $\mathrm{U}(\ensuremath{\nu}+1)$ group}, Phys.
  Rev. C 61 (2000) 067305.
\newblock \href {http://dx.doi.org/10.1103/PhysRevC.61.067305}
  {\path{doi:10.1103/PhysRevC.61.067305}}.
\newline\urlprefix\url{https://link.aps.org/doi/10.1103/PhysRevC.61.067305}

\bibitem{BIJKER2002334}
R.~Bijker, F.~Iachello,
  \href{https://www.sciencedirect.com/science/article/pii/S000349160296255X}{The
  algebraic cluster model: Three-body clusters}, Annals of Physics 298~(2)
  (2002) 334--360.
\newblock \href {http://dx.doi.org/https://doi.org/10.1006/aphy.2002.6255}
  {\path{doi:https://doi.org/10.1006/aphy.2002.6255}}.
\newline\urlprefix\url{https://www.sciencedirect.com/science/article/pii/S000349160296255X}

\bibitem{PhysRevLett.113.012502}
D.~J. Mar\'{\i}n-L\'ambarri, R.~Bijker, M.~Freer, M.~Gai, T.~Kokalova, D.~J.
  Parker, C.~Wheldon,
  \href{https://link.aps.org/doi/10.1103/PhysRevLett.113.012502}{Evidence for
  triangular $\mathcal{D}_{3h}$ symmetry in $^{12}\mathrm{C}$}, Phys. Rev.
  Lett. 113 (2014) 012502.
\newblock \href {http://dx.doi.org/10.1103/PhysRevLett.113.012502}
  {\path{doi:10.1103/PhysRevLett.113.012502}}.
\newline\urlprefix\url{https://link.aps.org/doi/10.1103/PhysRevLett.113.012502}

\bibitem{PhysRevLett.98.032501}
M.~Chernykh, H.~Feldmeier, T.~Neff, P.~von Neumann-Cosel, A.~Richter,
  \href{https://link.aps.org/doi/10.1103/PhysRevLett.98.032501}{Structure of
  the {H}oyle state in $^{12}\mathrm{C}$}, Phys. Rev. Lett. 98 (2007) 032501.
\newblock \href {http://dx.doi.org/10.1103/PhysRevLett.98.032501}
  {\path{doi:10.1103/PhysRevLett.98.032501}}.
\newline\urlprefix\url{https://link.aps.org/doi/10.1103/PhysRevLett.98.032501}

\bibitem{PhysRevC.81.024303}
S.~Hyldegaard, M.~Alcorta, B.~Bastin, M.~J.~G. Borge, R.~Boutami,
  S.~Brandenburg, J.~B\"uscher, P.~Dendooven, C.~{\relax Aa}. Diget,
  P.~Van~Duppen, T.~Eronen, S.~P. Fox, L.~M. Fraile, B.~R. Fulton, H.~O.~U.
  Fynbo, J.~Huikari, M.~Huyse, H.~B. Jeppesen, A.~S. Jokinen, B.~Jonson,
  K.~Jungmann, A.~Kankainen, O.~S. Kirsebom, M.~Madurga, I.~Moore, A.~Nieminen,
  T.~Nilsson, G.~Nyman, G.~J.~G. Onderwater, H.~Penttil\"a, K.~Per\"aj\"arvi,
  R.~Raabe, K.~Riisager, S.~Rinta-Antila, A.~Rogachevskiy, A.~Saastamoinen,
  M.~Sohani, O.~Tengblad, E.~Traykov, Y.~Wang, K.~Wilhelmsen, H.~W. Wilschut,
  J.~\"Ayst\"o,
  \href{https://link.aps.org/doi/10.1103/PhysRevC.81.024303}{$r$-matrix
  analysis of the $\ensuremath{\beta}$ decays of $^{12}\mathrm{N}$ and
  $^{12}\mathrm{B}$}, Phys. Rev. C 81 (2010) 024303.
\newblock \href {http://dx.doi.org/10.1103/PhysRevC.81.024303}
  {\path{doi:10.1103/PhysRevC.81.024303}}.
\newline\urlprefix\url{https://link.aps.org/doi/10.1103/PhysRevC.81.024303}

\bibitem{PhysRevC.93.054307}
Y.~Kanada-En'yo,
  \href{https://link.aps.org/doi/10.1103/PhysRevC.93.054307}{Isoscalar monopole
  and dipole excitations of cluster states and giant resonances in
  $^{12}\mathrm{C}$}, Phys. Rev. C 93 (2016) 054307.
\newblock \href {http://dx.doi.org/10.1103/PhysRevC.93.054307}
  {\path{doi:10.1103/PhysRevC.93.054307}}.
\newline\urlprefix\url{https://link.aps.org/doi/10.1103/PhysRevC.93.054307}

\bibitem{10.1093/ptep/ptw178}
Y.~Yuta, K.-E. Yoshiko, \href{https://doi.org/10.1093/ptep/ptw178}{{3$\alpha$
  cluster structure and monopole transition in $^{12}\mathrm{C}$ and
  $^{14}\mathrm{C}$}}, Progress of Theoretical and Experimental Physics
  2016~(12), 123D04.
\newblock \href {http://dx.doi.org/10.1093/ptep/ptw178}
  {\path{doi:10.1093/ptep/ptw178}}.
\newline\urlprefix\url{https://doi.org/10.1093/ptep/ptw178}

\bibitem{li2020multiprobe}
K.~C.~W. Li, F.~D. Smit, P.~Adsley, R.~Neveling, P.~Papka, E.~Nikolskii, J.~W.
  Brümmer, L.~M. Donaldson, M.~Freer, M.~N. Harakeh, F.~Nemulodi, L.~Pellegri,
  V.~Pesudo, M.~Wiedeking, E.~Z. Buthelezi, V.~Chudoba, S.~V. F\"ortsch,
  P.~Jones, M.~Kamil, J.~P. Mira, G.~G. O'Neill, E.~Sideras-Haddad, B.~Singh,
  G.~F. Steyn, J.~A. Swartz, I.~T. Usman, J.~J. van Zyl, Multi-probe study of
  excited states in $\mathrm{^{12}C}$: disentangling the sources of monopole
  strength between the {Hoyle} state and {$E_{x} = 13$ MeV} (2020).
\newblock \href {http://arxiv.org/abs/2011.10112} {\path{arXiv:2011.10112}}.

\bibitem{GargU}
U.~Garg, G.~Col{\`o},
  \href{http://www.sciencedirect.com/science/article/pii/S0146641018300322}{The
  compression-mode giant resonances and nuclear incompressibility}, Progress in
  Particle and Nuclear Physics 101 (2018) 55--95.
\newblock \href {http://dx.doi.org/https://doi.org/10.1016/j.ppnp.2018.03.001}
  {\path{doi:https://doi.org/10.1016/j.ppnp.2018.03.001}}.
\newline\urlprefix\url{http://www.sciencedirect.com/science/article/pii/S0146641018300322}

\bibitem{Harakeh_MN_vanDerWoude_A}
M.~N. Harakeh, A.~van~der Woude, Giant Resonances: Fundamental High-Frequency
  Modes of Nuclear Excitation, Oxford University Press, Oxford, 2001.

\bibitem{PhysRevC.83.037302}
B.~Mouginot, E.~Khan, R.~Neveling, F.~Azaiez, E.~Z. Buthelezi, S.~V. F\"ortsch,
  S.~Franchoo, H.~Fujita, J.~Mabiala, J.~P. Mira, P.~Papka, A.~Ramus, J.~A.
  Scarpaci, F.~D. Smit, I.~Stefan, J.~A. Swartz, I.~Usman,
  \href{https://link.aps.org/doi/10.1103/PhysRevC.83.037302}{Search for the
  giant pairing vibration through ($p$,$t$) reactions around 50 and 60
  {M}e{V}}, Phys. Rev. C 83 (2011) 037302.
\newblock \href {http://dx.doi.org/10.1103/PhysRevC.83.037302}
  {\path{doi:10.1103/PhysRevC.83.037302}}.
\newline\urlprefix\url{https://link.aps.org/doi/10.1103/PhysRevC.83.037302}

\bibitem{NEVELING201129}
R.~Neveling, H.~Fujita, F.~D. Smit, T.~Adachi, G.~P.~A. Berg, E.~Z. Buthelezi,
  J.~Carter, J.~L. Conradie, M.~Couder, R.~W. Fearick, S.~V. F{\"o}rtsch, D.~T.
  Fourie, Y.~Fujita, J.~G{\"o}rres, K.~Hatanaka, M.~Jingo, A.~M. Krumbholz,
  C.~O. Kureba, J.~P. Mira, S.~H.~T. Murray, P.~von Neumann-Cosel, S.~O'Brien,
  P.~Papka, I.~Poltoratska, A.~Richter, E.~Sideras-Haddad, J.~A. Swartz,
  A.~Tamii, I.~T. Usman, J.~J. van Zyl,
  \href{http://www.sciencedirect.com/science/article/pii/S0168900211012460}{High
  energy-resolution zero-degree facility for light-ion scattering and reactions
  at i{T}hemba {LABS}}, Nuclear Instruments and Methods in Physics Research
  Section A: Accelerators, Spectrometers, Detectors and Associated Equipment
  654~(1) (2011) 29--39.
\newblock \href {http://dx.doi.org/https://doi.org/10.1016/j.nima.2011.06.077}
  {\path{doi:https://doi.org/10.1016/j.nima.2011.06.077}}.
\newline\urlprefix\url{http://www.sciencedirect.com/science/article/pii/S0168900211012460}

\bibitem{Adsley:2016blb}
P.~Adsley, R.~Neveling, P.~Papka, Z.~Dyers, J.~W. Br{\"u}mmer, C.~{\relax Aa}.
  Diget, N.~J. Hubbard, K.~C.~W. Li, A.~Long, D.~J. Marin-Lambarri,
  L.~Pellegri, V.~Pesudo, L.~C. Pool, F.~D. Smit, S.~Triambak, {CAKE}: the
  coincidence array for {K}600 experiments, Journal of Instrumentation 12~(02)
  (2017) T02004--T02004.
\newblock \href {http://dx.doi.org/10.1088/1748-0221/12/02/t02004}
  {\path{doi:10.1088/1748-0221/12/02/t02004}}.

\bibitem{Konishi:2007:ICS:1554702}
S.~Konishi, G.~Kitagawa, Information Criteria and Statistical Modeling, 1st
  Edition, Springer Publishing Company, Incorporated, 2007.

\bibitem{Cavanaugh_1997}
J.~E. Cavanaugh,
  \href{http://www.sciencedirect.com/science/article/pii/S0167715296001289}{Unifying
  the derivations for the akaike and corrected akaike information criteria},
  Statistics \& Probability Letters 33~(2) (1997) 201--208.
\newblock \href
  {http://dx.doi.org/https://doi.org/10.1016/S0167-7152(96)00128-9}
  {\path{doi:https://doi.org/10.1016/S0167-7152(96)00128-9}}.
\newline\urlprefix\url{http://www.sciencedirect.com/science/article/pii/S0167715296001289}

\bibitem{CHUCK3}
P.~Kunz, \textsc{CHUCK3}, Coupled Channels Programme.

\bibitem{RevModPhys.90.035004}
M.~Freer, H.~Horiuchi, Y.~Kanada-En'yo, D.~Lee, U.-G. Mei\ss{}ner,
  \href{https://link.aps.org/doi/10.1103/RevModPhys.90.035004}{Microscopic
  clustering in light nuclei}, Rev. Mod. Phys. 90 (2018) 035004.
\newblock \href {http://dx.doi.org/10.1103/RevModPhys.90.035004}
  {\path{doi:10.1103/RevModPhys.90.035004}}.
\newline\urlprefix\url{https://link.aps.org/doi/10.1103/RevModPhys.90.035004}

\bibitem{PhysRevLett.116.132501}
M.~Scheck, S.~Mishev, V.~Y. Ponomarev, R.~Chapman, L.~P. Gaffney, E.~T. Gregor,
  N.~Pietralla, P.~Spagnoletti, D.~Savran, G.~S. Simpson,
  \href{https://link.aps.org/doi/10.1103/PhysRevLett.116.132501}{Investigating
  the pygmy dipole resonance using $\ensuremath{\beta}$ decay}, Phys. Rev.
  Lett. 116 (2016) 132501.
\newblock \href {http://dx.doi.org/10.1103/PhysRevLett.116.132501}
  {\path{doi:10.1103/PhysRevLett.116.132501}}.
\newline\urlprefix\url{https://link.aps.org/doi/10.1103/PhysRevLett.116.132501}

\bibitem{PhysRevC.86.024312}
Y.~Shimbara, Y.~Fujita, T.~Adachi, G.~P.~A. Berg, H.~Fujimura, H.~Fujita,
  K.~Fujita, K.~Hara, K.~Y. Hara, K.~Hatanaka, J.~Kamiya, K.~Katori,
  T.~Kawabata, K.~Nakanishi, G.~Mart\`{\i}nez-Pinedo, N.~Sakamoto, Y.~Sakemi,
  Y.~Shimizu, Y.~Tameshige, M.~Uchida, M.~Yoshifuku, M.~Yosoi,
  \href{https://link.aps.org/doi/10.1103/PhysRevC.86.024312}{High-resolution
  study of {G}amow-{T}eller transitions with the
  ${}^{37}${Cl}(${}^{3}\text{He},t$)${}^{37}${Ar} reaction}, Phys. Rev. C 86
  (2012) 024312.
\newblock \href {http://dx.doi.org/10.1103/PhysRevC.86.024312}
  {\path{doi:10.1103/PhysRevC.86.024312}}.
\newline\urlprefix\url{https://link.aps.org/doi/10.1103/PhysRevC.86.024312}

\bibitem{RevModPhys.30.257}
A.~M. Lane, R.~G. Thomas,
  \href{http://link.aps.org/doi/10.1103/RevModPhys.30.257}{R-matrix theory of
  nuclear reactions}, Rev. Mod. Phys. 30 (1958) 257--353.
\newblock \href {http://dx.doi.org/10.1103/RevModPhys.30.257}
  {\path{doi:10.1103/RevModPhys.30.257}}.
\newline\urlprefix\url{http://link.aps.org/doi/10.1103/RevModPhys.30.257}

\bibitem{RSmith_2019}
R.~Smith, J.~Bishop, C.~Wheldon, T.~Kokalova,
  \href{https://doi.org/10.1088%2F1742-6596%2F1308%2F1%2F012021}{Theoretical
  approaches to the 3$\alpha$ break-up of $^{12}\mathrm{C}$}, Journal of
  Physics: Conference Series 1308 (2019) 012021.
\newblock \href {http://dx.doi.org/10.1088/1742-6596/1308/1/012021}
  {\path{doi:10.1088/1742-6596/1308/1/012021}}.
\newline\urlprefix\url{https://doi.org/10.1088%2F1742-6596%2F1308%2F1%2F012021}

\bibitem{Pruet_2005}
J.~Pruet, S.~E. Woosley, R.~Buras, H.-T. Janka, R.~D. Hoffman,
  \href{https://doi.org/10.1086/428281}{Nucleosynthesis in the hot convective
  bubble in core-collapse supernovae}, The Astrophysical Journal 623~(1) (2005)
  325--336.
\newblock \href {http://dx.doi.org/10.1086/428281} {\path{doi:10.1086/428281}}.
\newline\urlprefix\url{https://doi.org/10.1086/428281}

\bibitem{Frohlich_2006}
C.~Fr{\"{o}}hlich, P.~Hauser, M.~Liebend{\"{o}}rfer, G.~Martinez-Pinedo, F.-K.
  Thielemann, E.~Bravo, N.~T. Zinner, W.~R. Hix, K.~Langanke, A.~Mezzacappa,
  K.~Nomoto, \href{https://doi.org/10.1086/498224}{Composition of the innermost
  core-collapse supernova ejecta}, The Astrophysical Journal 637~(1) (2006)
  415--426.
\newblock \href {http://dx.doi.org/10.1086/498224} {\path{doi:10.1086/498224}}.
\newline\urlprefix\url{https://doi.org/10.1086/498224}

\end{thebibliography}

\end{document}